\documentclass[preprint,aps,prd]{revtex4}
\usepackage{graphicx}
\usepackage{epsfig}
\begin{document}
\def\be{\begin{equation}}
\def\ee{\end{equation}}
\def\bearr{\begin{eqnarray}}
\def\eearr{\end{eqnarray}}
\def\tc{$T_c~$}
\def\tcl{$T_c^{1*}~$}
\def\c2{ CuO$_2~$}
\def\ruo{ RuO$_2~$}
\def\lsco{LSCO~}
\def\bi{bI-2201~}
\def\tl{Tl-2201~}
\def\hg{Hg-1201~}
\def\sro{$\rm{Sr_2 Ru O_4}$~}
\def\rc{$RuSr_2Gd Cu_2 O_8$~}
\def\mgb{$MgB_2$~}
\def\pz{$p_z$~}
\def\ppi{$p\pi$~}
\def\sqo{$S(q,\omega)$~}
\def\tperp{$t_{\perp}$~}
\def\he4{${\rm {}^4He}$~}
\def\ags{${\rm Ag_5 Pb_2O_6}$~}
\def\nxcob{$\rm{Na_x CoO_2.yH_2O}$~}
\def\lsco{$\rm{La_{2-x}Sr_xCuO_4}$~}
\def\lco{$\rm{La_2CuO_4}$~}
\def\lbco{$\rm{La_{2-x}Ba_x CuO_4}$~}
\def\half{$\frac{1}{2}$~}
\def\thalf{$\frac{3}{2}$~}
\def\tst{${\rm T^*$~}}
\def\tch{${\rm T_{ch}$~}}
\def\jeff{${\rm J_{eff}$~}}
\def\nbc{${\rm LuNi_2B_2C}$~}
\def\cabc{${\rm CaB_2C_2}$~}
\def\nboo{${\rm NbO_2}$~}
\def\voo{${\rm VO_2}$~}
\def\nip{$\rm LaONiP$~}
\def\nisb{$\rm LaONiSb$~}
\def\nibi{$\rm LaONiBi$~}
\def\fep{$\rm LaOFeP$~}
\def\cop{$\rm LaOCoP$~}
\def\mnp{$\rm LaOMnP$~}
\def\fesb{$\rm LaOFeSb$~}
\def\febi{$\rm LaOFeBi$~}
\def\efeas{$\rm LaO_{1-x}F_xFeAs$~}
\def\hfeas{$\rm La_{1-x}Sr_xOFeAs$~}
\def\hSfeas{$\rm Sm_{1-x}Sr_xOFeAs$~}
\def\hCefeas{$\rm Ce_{1-x}Sr_xOFeAs$~}
\def\feas{$\rm LaOFeAs$~}
\def\Ndfeas{$\rm NdOFeAs$~}
\def\Smfeas{$\rm SmOFeAs$~}
\def\Prfeas{$\rm PrOFeAs$~}
\def\refeas{$\rm REOFeAs$~}
\def\refesb{$\rm REOFeSb$~}
\def\refebi{$\rm REOFeBi$~}
\def\ttog{$\rm t_{2g}$~}
\def\fese{$\rm FeSe$~}
\def\fete{$\rm FeTe$~}
\def\eg{$\rm e_{g}$~}
\def\dxy{$\rm d_{xy}$~}
\def\dzx{$\rm d_{zx}$~}
\def\dzy{$\rm d_{zy}$~}
\def\dxsq{$\rm d_{x^{2}-y^{2}}$~}
\def\dzsq{$\rm d_{z^{2}}$~}
\def\LAO{$\rm LaAlO_3$~}
\def\STO{$\rm SrTiO_3$~}
\def\hsm{$\rm H_2 S$~}
\def\hm{$\rm H_2$~}

\title{My Random Walks in Anderson's Garden $^*$}

\author{ G. Baskaran}
\affiliation
{The Institute of Mathematical Sciences, C.I.T. Campus, Chennai 600 113, India \&\\
Perimeter Institute for Theoretical Physics, Waterloo, ON, N2L 2Y6 Canada}

\begin{abstract}

\emph{Anderson's Garden} is a drawing presented to Philip W. Anderson on the eve of his 60th birthday celebration, in 1983. This cartoon (Fig. 1), whose author is unknown, succinctly depicts some of Anderson’s pre-1983 works, as a blooming garden. As an avid reader of Anderson’s papers, a random walk in Anderson’s garden had become a part of my routine since graduate school days. This was of immense help and prepared me for a wonderful collaboration with the gardener himself, on the resonating valence bond (RVB) theory of High Tc cuprates and quantum spin liquids, at Princeton. The result was bountiful - the first (RVB mean field) theory for i) quantum spin liquids, ii) emergent fermi surface in Mott insulators and iii) superconductivity in doped Mott insulators. Beyond mean field theory - i) emergent gauge fields, ii) Ginzburg Landau theory with RVB gauge fields, iii) prediction of superconducting dome, iv) an early identification and study of a non-fermi liquid normal state of cuprates and so on. Here I narrate this story, years of my gardening attempts and end with a brief summary of my theoretical efforts to extend RVB theory of superconductivity to encompass the recently observed very high Tc $\sim$ 203 K superconductivity in molecular solid H$_2$S at high pressures $\sim$ 200 GPa. 

\vskip 2.0 cm
\noindent
$^*$ Closely follows an article published in \\ \textbf{PWA90~~A Life Time of Emergence}\\
\textbf{Editors:} P. Chandra, P. Coleman, G. Kotliar, P. Ong, D.L. Stein and Clare Yu\\ World Scientific 2016
\end{abstract}

\maketitle

\vskip 0.75 cm
\noindent {\bf{\large Introduction}}
\vskip 0.5 cm
I visited Phil Anderson at Princeton over the three-year period 1984-87.  An intense collaboration with Anderson, rather a resonance took place for few months,  November 1986 to March 1987.  I was drawn deep into the world of strongly correlated electron systems, novel quantum phases in Mott insulators and high Tc superconductors \cite{BednorzMuller,PWA1987}. Interestingly, while at Princeton, before high Tc cuprates appeared in the scene, I dabbled seriously with the idea of changing field into grey matter (neuroscience), an idea which Anderson encouraged. However, an enticing influence of new challenges from cuprates and Anderson's whole hearted involvement changed my course.

The first part of the present article is a personal account of my enjoyable stay at Princeton, collaboration with Anderson and a brief account of how I got involved in the theory of quantum spin liquids in Mott insulators and high Tc superconductivity in cuprates.  \emph{Anderson's Garden} is a cartoon drawing presented to him on the eve of his 60th birth day, in 1983, by a colleague (artist unknown ? \cite{AndersonGardenPicture}). This cartoon (figure 1) depicts a thoughtful Anderson and his earlier works.  From my Ph.D. days, random walk in Anderson's garden was my routine. Being an avid reader of Anderson's works was of immense help and prepared me for a wonderful collaboration with Anderson. 

\begin{figure}
\includegraphics[width=1.0\textwidth]{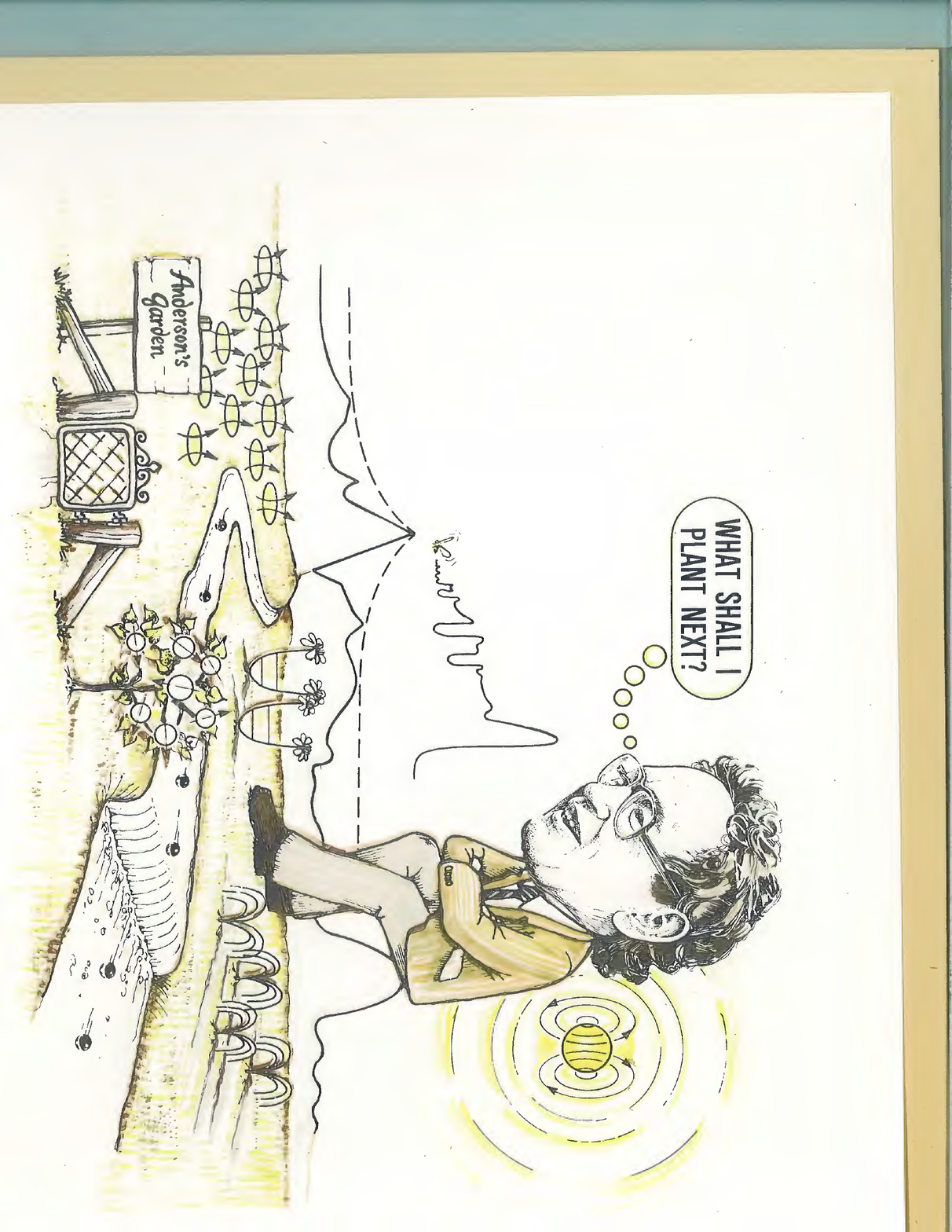}
\caption{Anderson's Garden: a sketch presented to Anderson by a colleague (name unknown \cite{AndersonGardenPicture}) at Bell Labs during the 60th birth day celebration} \label{fig1}
\end{figure}

Second part summarises my subsequent activities. Over years we have suggested electron correlations and RVB physics \cite{PWA1987} to be present \emph{to varying degrees}, in normal and superconducting phases of most new superconducting systems: fullerites, nickel borocarbides, hydrated sodium cobalt oxide, MgB$_2$,  ET and Bechgard organic family, boron doped diamond, iron arsenide family, doped graphene, doped TiSe$_2$, spin ladder compound and recently, doped silicene and germanene. I also predicted possibility of a p-wave superconductivity in Sr$_2$RuO$_4$, independent of Rice and Sigrist, by combining strong correlation and Hund coupling effects; and recently (with Gu and Jiang) possibility of chiral p-wave superconductivity in a 2-dimensional Nagaoka ferromagnet. Even family of \textit{doped band insulators} such as LaOBiS$_2$, in my view, create coulomb force driven self organized Mott insulators (a form of generalized Wigner crystals) and a rich superconducting scenario.

The third part of this article summarizes my work on the theory of superconductivity, discovered recently \cite{EremetsH2S} in molecular solid \hsm with a very high Tc $\sim$ 203 K, under a very high pressure of 200 GPa.  In \hsm molecule, four valence electrons form two saturated covalent bonds, H-S-H, and bind an S atom to two H atoms. We view paired valence electrons as \emph{confined Cooper pairs} and molecular solid \hsm as a \textit{Cooper pair insulator}. Pressure changes crystal structure, changes pattern of valence electron pairing and \textit{deconfines some Cooper pairs, before it liberates single electrons}. That is, i) sulphur atoms form a sublattice with saturated S-S covalent bonds, ii) part of H atoms left behind in the small interstitials of the sulphur subsystem forms a dilated H-atom sublattice, a Mott insulator with unsaturated H-H covalent bonds in a resonating valence bond state and iii) charge transfer between S and H subsystem, arising from a differing electro negativity  dopes the Mott insulator and leads  to superconductivity.

\vskip 0.75 cm
\noindent {\bf{\large 1. 1983-1984: Trieste to Princeton}}
\vskip 0.5 cm
My first meeting with Phil Anderson was at the International Center for Theoretical Physics, Trieste, Italy in the summer of 1983. I was in my mid 30's, visiting ICTP and SISSA for an extended period, after failing to get a permanent academic job in my home country. In those days, for many of us, theoretical physcists from third world countries, ICTP was a haven and played an important role in shaping our academic future. Now, after nearly 3 decades, theoretical physics scenario in India has improved in India, but much more is desired.

Erio Tosatti, my wonderful host at Trieste, had invited Phil Anderson for a colloquium. After the colloquium was over, Erio came rushing.  He said, `Phil is free, come and talk to him'. Even though I had a great admiration for Anderson, I was reluctant and some what shy to meet him, because of his stature in the field. However, Erio insisted that I meet. I agreed, after Arun Jayannavar, a good friend visiting ICTP, agreed to accompany me. 

The post lunch discussion with Phil lasted for more than an hour. Mostly I spoke. Anderson was in a sleepy/dreamy state; he made occasional remarks. I was describing my foray into CDW states, why supersolid $^4$He should exist in spite of a \textit{no-go theorem} of Anderson sketched in his book \textit{Basic Notions in Condensed Matter Physics}, my variational approach to quantum roughening in solid $^4$He and a few other topics that I was thinking about at that time. We parted. I was elated.

A continuing low job prospects back home and Erio's strong advice to cross the big ocean, forced me to try for research-cum-teaching visiting positions at the USA. I remember very well: at the end of a car ride near Trieste train station, Erio jotted down 13 names in a piece of paper. I wrote to all. Every one responded. Eleven negative and two positive responses. A warm response from Robert Schrieffer, whom I had met at ICTP on two occasions earlier, did not satisfy my time frame. Anderson responded positively. Actually he was apologetic that he could not fix a position for me at Bell Labs, as he had recently moved from Bell to Princeton. He wondered if I would accept a Visiting Research Staff (a visiting Assistant Professor position from Anderson's soft money) position at the Physics Department of Princeton University. I was overjoyed.

In early September of 1984, I landed at Princeton with my wife and 3 little children. Later I learned from Erio that Anderson enjoyed my post lunch discussion at ICTP.  From Anderson I learned that Erio recommended me strongly. Erio, a good friend and admirer of Anderson was his earlier Post Doc at Cambridge.

\vskip 0.75 cm
\noindent {\bf{\large 2. Random Walks in Anderson's Garden}}
\vskip 0.5 cm
\textit{Anderson's Garden} is a wonderful drawing, (figure 1) presented to Anderson on his 60th birthday, where Anderson's work is depicted as a nice garden. Gardener Anderson is wondering, resting on an Anderson localized state, `what shall I plant next ?'.  It is a very imaginative picture with a Neutron star (Alpar-Anderson-Pines-Sham vortex creep theory of neutron star quakes as origin of pulsar glitches ?) in the background, a buzzing bee depicting theory of motional narrowing phenomenon in magnetic resonance and so on. 

My random walk in Anderson's Garden started during my Ph.D. days, as a toddler. I was introduced to Anderson's works and helped to go into some depths by my friend Rajaram Nityananda, mentor Professor N Kumar, supervisor Professor K P  Sinha and mentor (late) Professor S K  Rangarajan.  Anderson localization theory, in the hands of Rajaram Nityananda and N Kumar, was exposed to us from different angles. Further, I ran a journal club - most talks were on Anderson's papers, as and when they appeared in the journals. For example, I reviewed Edwards-Anderson spin glass model as a series of papers were appearing. Random walks in Anderson Garden became a habit. 

Chandra Varma from Bell was lecturing at a TIFR summer school (1973) at IISc, Bangalore. He introduced us to Anderson lattice model, Hubbard model, Mott insulators, heavy fermions etc. Jayaraman of Bell Labs, who was on a sabbatical, setting up a high pressure laboratory at the National Aeronautical Laboratory at Bangalore also introduced us to the fascinating world of valence fluctuations, metal insulator transition in SmS, etc. Rajaram Nityananda, a fellow graduate student at that time, was ever ready to clear our doubts at the end of every lectures on any topic ! N Kumar, my mentor, got excited about topics in Mott transition, Jahn-Teller effect, Falicov-Kimbal model etc. I learned from C N R Rao, who had just joined IISc, fresh experimental results on the enigmatic Mott insulator LaCoO$_3$, that exhibited low to high spin cross over as a function of temperature, without any long range magnetic order. 

My supervisor K P  Sinha introduced me to Anderson's 1950 and 1959 papers on superexchange, in connection with a project he had suggested on electrical transport in doped EuO, a dilute magnetic semiconductor.  He had stories to tell us  about great physicists, including Phil Anderson, whom he met during his stint at Bell Labs. Rajaram Nityananda had explained to me, adding his own insights, Anderson’s Cargese lectures on local moment formation etc.  In 1977 H.R. Krishnamurthy arrived from Cornell and explained to us intricacies of Kondo phenomenon, valence fluctuation and how to understand them using quantum RG approach (built on Anderson's poor man's scaling theory) that Krishnamurthy, Wilkinson and Wilson had just developed.  Anderson's masterly role in modern condensed matter physics was manifest. Strong correlation physics, including Mott insulator, was in the air.

Other reason for a smooth entry to RVB theory at Princeton was my ealier interest in quantum magnetism and help from friends. I had studied consequences of (Jordan-Wigner) Fermi sea in 1D spin-half Heisenberg antiferromagnetic insulator and predicted in 1978, an unexpected Friedel type oscillation \cite{GBFriedelOscillations}; had used a projected wave function to go beyond spin-wave theory (an unpublished collaboration with Arun Jayannavar at ICTP) etc. Further I had exposure to i) Majumdar-Ghosh model and ordered valence bond states from Professor J. Pasupathy and my friend Sriram Shastry,  ii) anomalous S(q,$\omega$) and quantum dynamics in the critical 1D spin-half Heisenberg antiferromagnet from Shastry and  iii) Fazekas-Anderson work on RVB theory for 2D spin-half antiferromagnet on the triangular lattice from (late) Patrick Fazekas himself, at ICTP. I vividly remember Patrick explaining difficulties in fitting Marshall sign convention on a triangular lattice, and a consequent \emph{dynamic, floating} sign convention. It offered a possibility of viewing the spin liquid state as a quantum liquid of (topological) $\pi$-phase misfits. In the current parlance, it is a \textit{vison liquid} or Zheng-Yu Weng's \textit{phase string liquid} !

\vskip 0.75 cm
\noindent {\bf{\large 3. 1984-1987 @ Princeton: Cuprates Came at the End}}
\vskip 0.5 cm
My stay at Princeton was most enjoyable - talking to graduate students, Zhou Zou, Ted Hsu, Joe Wheatley, Yong Ren, Yaotian Fu, Mark Kwalay, Brad Marston; post docs Shoudan Liang, Anil Khurana, Ahmed Abouelsaood; faculty, Phuan Ong, Dan Stein, Jim Sauls, Sajeev John, Stuart Trugman, Ian Affleck; discussing with them on a variety of modern and old topics. Xiao-Gang Wen, then a string theory graduate student, was often a silent spectator in our condensed matter discussions. 

Discussion with Anderson was special. We provided a mutual sympathy field and talked on topics that went beyond condensed matter physics. For example, he used tell me how low energy nuclear physics is incomplete, if it uses only 2 body nuclear potentials; vacuum structure of QCD might have important role to play etc. We discussed whether human made machines will ever develop qualities such as \textit{self consciousness}. During my stay Anderson offered two graduate courses on theoretical biology. Having developed some interest in Biology during my Ph.D. days at Bangalore, and having an eye on neuroscience at Princeton, I enjoyed both courses. 

Beyond condensed matter physicists, being in the company of many legendary figures at Princeton was a great treat for me. Princeton University and Institute for Advanced Studies also attracted great minds from all over the world, regularly for colloquia, seminars and visits. 

On the family front, we had wonderful time at Princeton. We went on outings on most weekends; often strayed into states of New York, Pennsylvania and Washington DC; several long drives to Niagara falls. Also a memorable trip to Florida for 2 weeks  in a motorhome. I drove the motorhome, filled with 3 families (a total of 14 adults and children) in Indian style. We established life long friendship with wonderful people.

It was November 1986. And end of nearly two and a half years of stay at Princeton, from September 1984, without any substantial achievement, except for few papers on cooperative ring exchange theory of fractional quantum Hall effect, spin glasses and travelling salesman problem. I had started feeling uneasy about my non publishing mode. However, Anderson did not seem to mind. Once, at the beginning of my 2nd year of stay, he made a comforting and prescient remark: `To do anything substantial it will take 3 years'. He was kind. He seems to have appreciated my regular interaction with him, students, faculty, visitors and others.

It was early November 1986. One early morning, Anderson came to my office and placed a photocopy of an article \cite{BednorzMuller} on my table, and said, we should work on this. The article was, soon to become a Nobel Prize winning one by Bednorz and Muller, just published in Z. fur Phys. It reported discovery of high Tc-superconductivity in cuprates. A friend of Anderson (Ted Geballe ?) had alerted Phil about this discovery by Bednorz and Muller, after listening to a talk by  Kitazawa at Bell Labs. Kitazawa had introduced the exciting Bednorz-Muller discovery, confirmation by Tanaka's group (late Kitazawa was a key member of Tanaka's group) and happenings beyond. Anderson and I missed Kitazawa's talk. Anderson used to drive to Bell Labs to attend interesting talks. I have accompanied him some times. On one of these visits I had the pleasure of listening to a wonderful talk by Richard Feynman at Bell, in the fall of 1984. Feynman was far ahead of his time and talked about ways to enable communications among nanoscale qubits in a quantum device !

I read through the article of Bednorz and Muller carefully. Having accepted a job at the Institute of Mathematical Sciences, at Madras (now Chennai), India and having decided to join in mid 1987, I was left with only about half year of stay at Princeton in Anderson's group. A wonderful opportunity to collaborate with Anderson had opened up. Secondly, having been introduced to Jahn-Teller effects in oxides from my Ph.D. thesis advisor Prof K P Sinha and my mentor Prof N Kumar at the Indian Institute of Science, Bangalore, I got excited about Jahn-Teller bipolaron mechanism of high Tc superconductivity that Bednorz and Muller were hinting at. I sat down and developed a multi band model for high Tc superconductivity, incorporating Jahn Teller effect.

A week later I reported my  progress to Anderson. He listened carefully. Made a brief and frank remark, `\textit{your theory is beautiful, but is totally irrelevant for cuprate superconductivity}'.  My level of excitement about my own theory was high. \textit{It made me deaf to Anderson's critical remark}. I did not know what was in Anderson's mind at that time. From the beginning I knew that the parent compound \lco is a Mott insulator \cite{PWAPersonalHistory}, thanks to T P Radhakrishnan (a graduate student friend from Princeton Chemistry Department). Radhakrishnan gave me an important experimental paper by Ganguly and Rao \cite{GangulyRao} (late Ganguly, my friend and Professor CNR Rao, a well wisher from IISc, Bangalore), on the Mott insulating antiferromagnet \lco. Unlike Anderson, I was insensitive to the nearness of high Tc superconductivity to a Mott insulator, and kept going my way. I even invited my friend and collaborator Dung-Hai Lee from IBM, Yorktown Heights to Princeton to spend a week, to begin a collaboration on this important problem.  Incidentally, Dung-Hai and I had great time as collaborators. We used to talk for hours over the telephone. I also spent a few memorable summers at IBM York Town Heights, visiting Dung-Hai.

\vskip 0.75 cm
\noindent {\bf{\large 4. RVB Mean Field Theory, Spin Liquids, Emergent Fermi Surfaces and Superconductivity from Doped Mott Insulator}}
\vskip 0.5 cm
It was mid December 1986. Anderson was leaving Princeton for Bangalore, for a conference on valence fluctuations.  On the day before he was leaving, as a parting remark he told me, `Baskaran, the whole thing is a spin-\half Mott phenomena - think about it'. I am used to short but loaded one or two line remarks from Anderson. I took them seriously. In fact, I enjoyed thinking about them, like a puzzle. The present statement got registered in my mind and made some subconscious rumblings. However, I continued on the Jahn-Teller path. I was confident that I could convince Anderson about Jahn-Teller mechanism for cuprates and eagerly awaited his return to the US in January 1987. As planned, Anderson flew from India to California, to spent some time at Caltech.

I couldn't wait and called Anderson the day he arrived at Caltech, to discuss my progress. Phil was quick to divert my attention. He said, `Baskaran, I have seen the light'. He further added, `\emph{Our starting point is a Mott insulator in a resonating valence bond state. We are sitting on a cusp. Doping produces superconductivity}'.  These three sentences did magic to me. It was some kind of revelation. I told him somewhat hurriedly `I think I understand what you are saying. Will call you back shortly'.  

At that time Dung-Hai Lee and Zhou Zou were in my office. I elaborated to them Anderson's three sentences, adding my own interpretation of how phase coherence among valence bond configurations in the Mott insulating state could emerge as superconductivity after doping. After half hour I called Anderson and explained to him my understanding of his three sentences. He remarked happily, `you have smelled it right'.  That was the beginning of a most enjoyable, satisfying and continuing collaboration. In a week's time a manuscript on a mean field theory of quantum spin liquid and RVB mechanism of superconductivity \cite{BZA}, co-authored with Anderson and Zou, was ready. Zou was a very smart graduate student and a good collaborator. It was a loss for us as he left physics after spending years at Princeton, Stanfors and Institute for Advanced Studies.

Another happy coincidence. During my random walks in Anderson's garden in 1986, but several months before cuprates appeared in the scene, I read twice, Anderson's 1973 RVB theory paper \cite{PWA1973}. This is a key and fundamental paper on quantum spin liquids. I wondered about meaning of the \textit{phase coherence} among the valence (singlet) bonds in this insulating ground state. It also intrigued me that such phase coherence is simply absent in a normal band insulator, but present in a superconductor.  I dont' know why, but I read this particular paper more than once.  These visits to Anderson's garden prepared me for a collaboration, soon to happen.

Our paper with Zou and Anderson \cite{BZA} was the first theory of quantum spin liquids, using a physically motivated enlargement of Hilbert space and a mean field theory. We applied the mean field theory to undoped and doped Mott insulators. We focussed on the constituent electron degree of freedom, rather than local moments. We boldly worked on an enlarged Hilbert space and suggested that proper incorporation of phase fluctuation of RVB order parameter should bring us back to physical Hilbert space.  The idea of decoupling the spin-spin interaction term in terms of Cooper pair operator, came from a paper of Noga \cite{Noga} written in the context of Anderson lattice model.  We obtained a pseudo fermi surface and a quantum spin liquid in the Mott insulator with practically no effort. Soon slave particle methods \cite{SlaveParticle} and Gurzwiller approximation scheme \cite{GutzwillerApproximation} followed at the heels. 

How our first article got published is interesting in itself. If Anderson had a fear for any one in the field of physics, it was journal referees. I can very well imagine fear on the other side, ending up as a referee for Anderson's paper !  Anderson corrected our manuscript and send it back (faxed ?) to me, from Caltech. I submitted it to PRL, thinking of it as a natural destination. When I told him, Anderson was worried. He suggested an immediate withdrawal from PRL and submission to Solid State Communication.  His concern was a potential delay from referees, because ideas were new and some what radical. The irony was, in that exciting initial months following high Tc cuprate discovery,  papers on high Tc cuprates received by PRL were refereed by a panel of experts; instant decisions were made. Apparently PRL had already accepted our paper, by the time our withdrawal request reached them. We were unaware of the acceptance. Respecting our withdrawal request PRL obliged.  It is Solid State Communications's turn now. After some exchange with a referee and consequent delay, our paper got published in Solid State Communications.

\vskip 0.75 cm
\noindent {\bf{\large 5. Emergent Gauge Fields}}
\vskip 0.5 cm
At IISc, Bangalore, I had a very good course on quantum field theory by Professor R Rajaraman (author of the famous book, Solitons and Instantons). He was an outstanding teacher, like R. Shankar (his youngest brother and my friend) at Yale. Rajaraman made students feel at home with difficult concepts. Soon after this exposure, I listened to a lecture series on lattice gauge theory and renormalization group by Leo Kadanoff, another series  by Franz Wegner at a Summer School on Statistical Mechanics (1976) at Sitges, Spain. (ICTP supported my visit to Spain, from Trieste). This got me interested in non-Abelian lattice gauge theory, structure of QCD vacuum, strong coupling approach to glue ball masses etc.  

On Prof G Rajasekaran's invitation I joined the vibrant Department of Theoretical Physics of University of Madras, as a temporary reader, in late 1978. I gave a short course on lattice gauge theory in 1980, to students and colleagues who were all high energy physicists. Field theory on a lattice made things more transparent to me. Elitzur's theorem, on impossibility of spontaneous breaking in presence of a local gauge symmetry  became easy.  Being in a group of very active high energy physicists for nearly 4 years,  I got exposed to a variety of challenging high energy physics problems and quantum field theory issues.

The projective aspect, namely Hilbert space restriction resulting from removal of double occupancy (Gutzwiller projection) in the low energy description of Mott insulator and doped Mott insulator is of paramount importance in Anderson's starting point. It became clear to me that this Hilbert space restriction in the Mott insulator produces an emergent local U(1) symmetry, and dynamically generated gauge fields, when we describe physics in terms of the underlying physical electrons. I used Elitzur's theorem, a consequence of emergent local U(1) symmetry, to prove that ODLRO exhibited by our mean field theory at half filling is only an artefact and it can be easily removed. The phase fluctuations, not manifest in the spin language, captures spin singlet and spin dynamics, leading to dynamical gauge fields. Doping converts the local U(1) symmetry to a global one and allows for possibility of superconductivity. Elitzur theorem does not apply when we have a global symmetry. 

I communicated my gauge theory calculations to Anderson around March 1987. Phil and his wife Joyce were spending their usual spring break, a month of retreat at Cornwal, a coastal village in UK.  Anderson replied that he has come to similar conclusions and explained it. This second resonance and a chance for another collaboration got me even more excited. This is the origin of our paper \cite{BAGauge} `Gauge Theory of High Temperature Superconductors and Strongly Correlated Fermi Systems'. This is one of the beginnings of the currently popular emergent gauge fields in condensed matter systems.

Our free energy for the spin liquid state in a Mott insulator, in terms of the bond singlet variable, has a local U(1) symmetry.  Doping reduces the symmetry to a global U(1) symmetry. However,  our action, a generalized RVB Ginzburg-Landau free energy expression for doped Mott insulators,  \textit{had memory of the Mott insulator} in the superconducting state.  Interestingly, our free energy also had d-wave solution as the lowest energy solution. However, we were sticking to extended-s symmetry solution because of the then available experimental results that indicated pseudo fermi surface like behaviour in the superconducting state. \textit{To Anderson, experiments come first.} Our hypothesis was that strong U(1) gauge field fluctuations will stabilize an extended-s wave solution, as opposed to d-wave solution. This turned out to be not the case. The d-wave solution that Kotliar-Liu, Michael Ma and others found has stood the experimental test.
 
 In our paper we hinted at a hidden local particle-hole symmetry (Z$_2$), in addition to the local U(1) symmetry. According to Anderson \cite{PWAPersonalHistory} I told him about SU(2) local symmetry, while finishing our article. The U(1) and Z$_2$ got nicely woven into a beautiful SU(2) local gauge theory, in a formal way, by Anderson, Affleck, Zou and Hsu and independently by Dagotto, Fradkin and Adriana Moreo \cite{SU2GaugeTheory}. 
 
The community was quick to accept our idea and theory of emergent gauge fields. Very soon connection of the magnetic fluxes and electric fields of the emergent U(1) RVB gauge fields to spin current (chirality) (Wiegmann, Wen, Wilczek and Zee) and valence bond density (Read and Sachdev) were established \cite{WWZ}. Phenomenological and microscopic consequences of emergent gauge fields, for normal and superconducting properties of cuprates, were worked out by Ioffe-Larkin, Patrick Lee-Nagaosa, Paul Wiegmann \cite{NagaosaLeeWiegmann} and others in some key papers. Suggestion of a topological Hopf term and a consequent statistics transmulation, in the 2d spin-\half quantum antiferromagnet, in a gauge theory description by Dzhyaloshinski, Wiegmann and Polyakov \cite{DPW} also excited the community. 

\vskip 0.75 cm
\noindent {\bf{\large 6. Spin Charge Decoupling, Anomalous Metallic State Etc.}}
\vskip 0.5 cm
As we were understanding mechanism of superconductivity and developing approximation methods to study superconducting state, it became clear that Anderson's mind was getting diverted into the anomalous normal state of the optimally doped cuprate. Interestingly, Anderson was very satisfied with local superexchange as the pairing mechanism and origin of a strong pairing scale, our RVB mean field theory and several important notions and scenarios that emerged with it. 

The now well known Temperature-Doping phase diagram with a \textit{dome}, was predicted on very general grounds in our PRL in 1987 \cite{ABZH}, even before such a phase diagram was experimentally measured. 
An insightful prediction of spin charge decoupling by Kivelson, Rokhsar and Sethna \cite{KivelsonRokhsarSethna} was brought into some what sharp focus in the optimally doped metallic state in this paper. While Kivelson, Rokhsar and Sethna \cite{KivelsonRokhsarSethna} coined the name \textit{holon}, the unpaired neutral fermion that Anderson introduced in his 1987 paper remained nameless. Anderson christened it `spinon', in our paper \cite{ABZH}. 

Our paper \cite{ABZH} also talked about possible origin of linear resistivity in the (non-fermi liquid) normal state, based on a simple golden rule estimation of scattering of (incoherent) holons by the fermionic spinon quasi particle excitations at the pseudo fermi surface. Our paper also emphasized that doping of the Mott insulator does not produce a rigid displacement of underlying (spinon) fermi surface. Essentially, only part of the spinon fermi surface in k-space is carved out, while accommodating doped holes. Currently popular Fermi arcs and small fermi surfaces seen in the pseudo gap phase has its origin in this old insight.

While Anderson respected that a theory should be able to describe physics qualitatively and quantitatively, he also realized that a straight forward theoretical analysis is going to be tough, because of the projection and a consequent strong coupling character; one should not get intimidated by difficulties and and get help wherever it comes from, be it mean field theories or phenomenology. Similarly successive reduction and model building, using microscopics as well as phenomenology, is an important part of the game. They are physics and phenomenology motivated renormalization procedures. A model should be simple and not simpler (sic).

It also became clear to us that tJ model, introduced by Anderson to describe cuprate physics is more appropriate for the optimally doped region. Underdoped pseudo gap region is dominated by residual unscreened coulomb interactions, disorder effects and electron-phonon coupling, not contained in the tJ model. The best way to understand the mechanism of superconductivity in its purest form is in the optimally doped regime. Anderson used to call experimental phenomenon seen in under doped regime as arising from nanoscopic phase separation and metallurgical complications. 

However, thanks to continuing experimental efforts, new physics such as existence of `small fermi surface', hiding in the background of a variety of competing phases, has emerged from quantum oscillation experiments, for example. This is unexpected, as we have a strong memory of the Mott insulator in the form of very ill defined quasiparticles at the chemical potential as shown by ARPES. 

I have offered an explanation \cite{GB3/2FermiLiquid} for existence of a pseudo fermi liquid having a small fermi surface by applying the idea of Haldane exclusion statistics \cite{ExclusionStatistics} to the doped hole in a short range spin liquid reference state in a Mott insulator. That is, doped hole is a composite object, a loosely bound (by RVB gauge forces) topological excitations namely holon and a spinon, living in a reference neutral spin liquid state, rather than a band insulator. Holon and spinon have Haldane exclusion statistics of 1 and $\frac{1}{2}$, that adds up to an  exclusion statistics of $\frac{3}{2}$ for a hole. I have called this fermi liquid like state as a $\frac{3}{2}$ fermi liquid. A conventional hole, a fermion in a band insulator, has an exclusion statistic of 1 and occupies one elementary k-space cell. However, each 3/2 fermion occupies, on an average, 3/2 elementary k-space cell. This leads to an enlarged \textit{fermi momentum}, consistent with a $\sim$ 3/2 times expansion of Fermi pockets seen in quantum oscillation experiments \cite{QOscillations}. 

\vskip 0.75 cm
\noindent {\bf{\large 7. Are We Bees ?}}
\vskip 0.5 cm
I had a wonderful visit to Aspen in the summer of 1987. By that time I was convinced that the resonating valence bond states advocated by Pauling in the context of p-$\pi$ bonded molecules and graphite, and elevated to novel quantum spin liquid states in Mott insulators in 1973 by Anderson should be ubiquitous beyond cuprates. Anderson and I had talked about possible role of RVB physics, in doped BaBiO$_3$ and so called \emph{bad actors} A15 and Chevral phase superconductors. BaBiO$_3$ is interesting.  It was popularly known as a negative U Hubbard system because of valence skipping and an apparent charge disproportionation. However, an early analysis of spectroscopic results by Kasuya, did not support the charge disproportionation and negative U idea. This gave Anderson and me confidence to think about repulsive Hubbard model for doped BaBiO$_3$.

\begin{figure}
\includegraphics[width=1.35\textwidth, angle = 270]{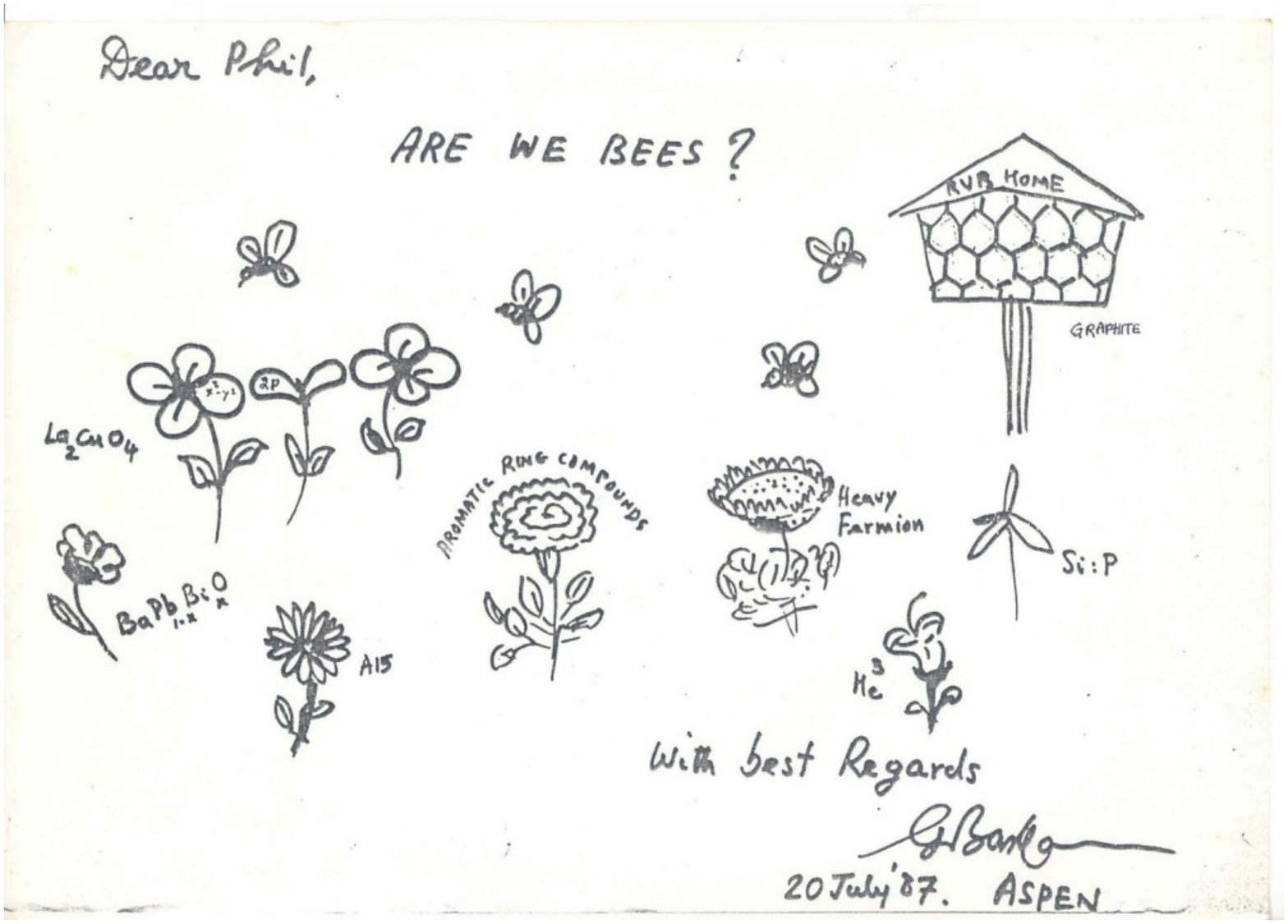}
\caption{A hand drawn post card sent to Anderson by the author in the summer of 1987 from Aspen \cite{PWA.AIP}. The first flower on the left with 4 petals is marked $x^2-y^2$ (3d$_{x^2-y^2}$ orbital) and that with 2 petals is marked 2p (2p$_x$ orbital).} \label{fig2}
\end{figure}

This thought, that there may be other systems with RVB physics was high in my mind at the Aspen Center for Physics. Aspen is a beautiful mountain resort and a great place to contemplate. There is music in the background, flowers, mountains and possibilities of nice hikes in the summer; small number of lectures and generally free, welcome and friendly atmosphere. Anderson, Elihu Abhrahams, Ravin Bhatt, Steve Kivelson, David Pines and several others had worked hard to maintain Aspen Center as a great Physics Center in the world.

Having been obsessed with cuprates, flower petals in the garden reminded me of d-orbitals of copper and 2p orbitals of oxygen in the CuO$_2$ layers. I took that opportunity to make a quick sketch and posted a card to Anderson (figure 2). Anderson liked my sketch and its contents and put it as a slide in one of his talks. This picture appears in one of the AIP Conference proceedings, with Anderson's comments \cite{PWA.AIP}. 

Looking back, this picture emphasized graphite (a single layer of which is graphene) as RVB home. In the wake of superconductivity in MgB$_2$ this idea surfaced again \cite{GBMgB2} and the result was my prediction of very high Tc in doped single sheet graphite, based on a tJ0 model, that is, a tJ model with no double occupancy constraint.  Graphite is not a Mott insulator; it is a highly anisotropic semi metal. However it has a strong nearest neighbour covalent bond or singlet pair correlation, according to Pauling. The tJ0 model I introduced thus combined band physics and valence bond (spin singlet) physics in a semi microscopic phenomenological, but microscopic fashion. To my great satisfaction, my work was pursued by good friend Seb Doniach and his student Annica Black Shaefer \cite{DoniachAnnica}. They discovered an unconventional d + id  chiral superconductivity solution, with a very high superconducting Tc. A variational Monte Carlo calculation \cite{PathakShenoyGB} with Vijay Shenoy and Sandeep Pathak, that went beyond mean field theory and took into account quantum fluctuations brought the scale of Tc down, for the chiral superconducting state. It was still high and a welcome value of 200 K ! There are also other parallel theoretical developments confirming the above \cite{NandKishoreThomaleZhengcheng}. Experiments have not confirmed our prediction, possibly because of an unavoidable disorder that comes at the desired range of doping and a high sensitivity of d + id state to disorder. 

I should also point out that experimental groups of Kopelevich and Esquinazi \cite{KopelevichEsquinazi} have reported unstable and tenuous signals for superconductivity, reaching room temperature scales in perturbed (doped) graphite. Prospects for high temperature superconductivity in doped graphene and sister compounds silicene and germanene \cite{GBSilicene} are there from theory point of view. It needs to be explored experimentally.

This picture (figure 2) has P doped Si, with 4 flower petals mimicking the sp$^3$ tetrahedron bonds. I had imagined a correlation based narrow phosphorous impurity bond superconductivity. Later experiments showed superconductivity, but the scale was too low below 1 K. Fortunately B doped diamond exhibited a higher Tc of 12 K later in 2004. I had attempted to explain this \cite{GBDiamond} as superconductivity in a self doped impurity band Mott insulator, occurring close to the Anderson-Mott transition point, along the doping axis. 

I dont' dare say I predicted superconducting C$_{60}$ compound, even though the big flower (with `aromatic ring compounds' written on top) was supposed to mimic a big molecule with p-$\pi$ bonds and ring currents.  BaBiO$_3$ and A15 compounds, that appear in the picture, once in a while surface from my subconscious mind, even now.

The message of this section is that \emph{Anderson inspires}.

\vskip 0.75 cm
\noindent {\bf{\large 8. Return to India: Gardening beyond Cuprates and a Synthesis}}
\vskip 0.5 cm
Having accepted a job at the Institute of Mathematical Sciences, Chennai I returned to India in late 1987. It was a difficult decision, as my collaboration with Anderson on high Tc superconductivity made me an instant hero, with tempting tenured job offers from outstanding places in the US and Europe. In a span of 7 months in 1987, after I began my work with Anderson on high Tc superconductivity, I gave nearly 35 talks all over north America, Japan and a few places in Europe.  As there were new concepts, notions and techniques, my talks often got stretched to several hours of discussions and working out details. The maximum duration was, 12 hours, at Hide Fukuyama's group in Tokyo in the summer of 1987. It was similar when I visited Vinay Amabagaokar at Cornell.  I was excited by the captive audience, wherever I went. Probably audience got enthused by my excitement.

Soon I became an \emph{RVB preacher}. ICTP, Trieste provided a podium. I got involved in various activities on strongly correlated electron systems and high Tc superconductivity on an yearly basis at ICTP, (1988 to $\sim$ 2005) thanks to invitation, appreciation and support from Abdus Salam, Erio Tosatti, Yu Lu, Mario Tosi, Norman March and Stig Lundquist. 

My friends suspected that I smelled RVB physics in any new superconductors that emerged in the scene. It started with K$_3$C$_{60}$, a fullerite. With Erio Tosatti we developed a mechanism for superconductivity \cite{GBTosattiK3C60} that used the valence bond correlations in the fullerene molecules and a consequent pair binding, a notion that was independently introduced by Kivelson and Chakravarty \cite{KivelsonChakravarty}. Anderson was very supportive of our theory.  Molecular conduction bands in K$_3$C$_{60}$ were very narrow, less than 0.25 eV. How a stoichiometric compound K$_3$C$_{60}$  manages to be a metal was already a surprise to Anderson. Mott localization should be imminent. I fondly remember a discussion on this issue between Phil, Walter Kohn and I. Later experiments by Iwasa and others showed that indeed superconducting K$_3$C$_{60}$ can be converted into a Mott insulator by a negative pressure, an expansion, by incorporating inert NH$_3$ molecules in interstitial sites in the unit cells of K$_3$C$_{60}$.

The next superconductor in line was the nickel borocarbide family, discovered by the TIFR group at Mumbai and Cava and others in the US. It was a layered system. Interestingly the layer structure is similar to the FeAs layers in the Fe based superconductors. I was convinced of a mechanism of superconductivity based on electron correlation \cite{GBNiBC}. Neutron scattering indicated a strong $(\pi,\pi)$ type magnetic fluctuations, and even a signal for a neutron resonance mode. In view of a complicated looking band structure, modelling was some what complicated and a deeper understanding remains obscure.

To me organic superconductors, often in the vicinity of a Mott insulator was a mystery for a long time. How pressure converts a Mott insulator into a superconductor ? I realized, from the existing phenomenology that the superconducting side of Mott transition point is better thought of as a  (lightly) self doped Mott insulator. The self generated small density of holons and doublons, of equal density, in the half filled band is determined by some kind of Madulung energy again, in line with an inevitable long range interaction on the Mott insulating side; but Hubbard model misses this. So I suggested a two species t-J model to understand superconductivity in organics \cite{GBOrganics}. This theory was of immense satisfaction to me, as it unified superconductivity in cuprates and organic superconductors.

Then came Na$_x$.CoO$_2$.yH$_2$O, a hydrated sodium cobalt oxide superconductor. Narrow band and correlation based physics superconductivity was obvious. I predicted a d + id chiral RVB type of superconductivity \cite{GBNaCoO}. Charge ordering in the CO$_2$ layer and ordering in the intercalant Na layer and role of H$_2$O molecule complicated the physics.

In my view, Fe arsenide superconductor is an example of double RVB system \cite{GBFeAs}, where two valence electrons in the 3d$^6$ shell of Fe$^{2+}$ form some kind of Hund coupled 2d spin half RVB system with internal charge transfer.  Unfortunately this family is also complex, unlike the cuprates, where a single band makes life much simpler. The cousin of Fe arsenide, FeSe and FeS seem to be making things simpler. A better understanding of superconductivity might emerge from these systems. There exists a vacancy ordered
system K$_2$Fe$_4$Se$_5$ with a (high) spin-2 Fe moments that exhibits an intriguing superconductivity, in the background of a long range antiferromagnetic order, involving a nearly spin-8 sublattice moment of a plaquette of Fe ions. I have suggested that this is a charge -2e skyrmion based 
superconductor \cite{GBSkyrmionSupCond}.

Silicene, Germanene and Stanene are 2d analogues of graphene. C,Si,Ge and Sn occur in the same column in the periodic table. However the atomic radii of Si and Ge are about 60 percent larger than that of carbon. According to electronic structure calculations this leads to a substantial,  3 fold reduction in the band width of p-$\pi$ band in silicene and germanene.  Based on band theory estimates of the t and U parameter and some overwhelming phenomenology I came to the conclusion that silicene and germanene are likely to be Mott insulators \cite{GBSilicene}. This is a prediction that is yet to be confirmed, because of not being able to synthesize free standing silicene or germanene on insulating substrates. The t and J parameter I estimate for silicene makes it a prospective playground for room temperature superconductor, provided competing phases such as valence bond ordering are kept under control.

Few other interesting systems of our interest are superconductivity in spin ladder compound \cite{GBSpinLadder} and a recently popular doped TiSe$_2$ \cite{GaneshGB} with potential chiral spin singlet superconductivity, in line with earlier doped graphene and hydrated cobalt oxide.  

Thanks to Piers Coleman's comments and provocation at a strong correlation workshop at ICTP Trieste, I ended up predicting \cite{GBPWave} p-wave superconductivity in Sr$_2$RuO$_4$, independently of Rice and Sigrist \cite{RiceSigrist}. Very interestingly, our recent study \cite{ZhengchengGB} of Infinite U Hubbard model in 2d honeycomb lattice, with Zhengcheng Gu and Hong-Chen Jiang, supports chiral p + ip superconductivity riding on on Nagaoka Ferromagnetism. It is a new twist to Nagaoka Ferromagnetism, perhaps universal.

A family of doped band insulators show intriguing superconductivity, that resembles doped Mott insulating cuprates. Recently I studied LaOBiS$_2$ \cite{GBLaOBiS2}, a prototype band insulator belonging to this category. The dilute density of doped carriers, make use of orbital degeneracy and residual long range coulomb interaction and self organize emergent Mott insulators, in the form of generalized Wigner crystals. This leads to the possibility of RVB superconductivity, in a doped band insulator \cite{GBLaOBiS2}.

Based on my experience with cuprates and later works indicated above, I suggested a \emph{5 Fold Way to New Superconductors} \cite{GB5FoldWay}, a guideline for experimental colleagues. Here electron correlation and relatively narrow bands play a central role. The five routes are: i) Copper Route (doped spin-half Mott insulator), 2) Pressure Route (Self Doped Mott Insulator in Organics), 3) Diamond Route (RVB physics in impurity band), 4) Graphene Route (Broad band 2d and intermediate correlations) and 5) Double RVB Route (Fe Arsenide and self doped spin-1 Mott insulators).

\vskip 0.75 cm
\noindent {\bf{\large 9. {Regular Visits to Princeton and Working with the Gardener}}}
\vskip 0.5 cm
I have been visiting Princeton regularly, on yearly basis and continuing my collaboration with Phil Anderson, ever since I returned to India in late 1987. Thanks to Anderson's hospitality and hospitality of friends like Phuan Ong, Duncan Haldane, Shivaji Sondhi and several others. these visits are always like returning home.
On the first day of my visits I will have a long chat with Anderson and summarise my year of activities. Anderson's quick grasp was phenomenal. To my delight, often Anderson's response was a nod of approval.  I will be soon drawn by Anderson into what excited him most in those days. It was often on cuprate physics. I will also have extended discussions with Phuon Ong and catchup with his regular and exciting experimental discoveries.

I had a second extended visit to Princeton, during 1995-1996, as a Member of the Institute for Advanced Studies, followed by a visit to Physics Department of Princeton University. At the Institute Frank Wilczek was a wonderful host. This visit was also marked by my participation in a remarkable course on molecular biology meant for physicists, organized by Stan Leibler. My class fellows were Frank Wilczek and Curt Callan among others ! This course was very effective, as each participating physicist was paired up with a graduate student from biology, who acted as a kind of tutor for us.

All these visits have been enriching, Random walks in Anderson's garden, often working together with the wise gardener continued. My latest visit to Anderson, in August 2015, also turned out to be as enjoyable, as ever.

\vskip 2.75 cm
\noindent {\bf{\large 10. Theory of Cooper Pair Crystals and Superconductivity in Molecular Crystals Under Pressure}}
\vskip 0.5 cm

Solid hydrogen has evoked a great interest over decades, from condensed matter to planetary physics community. Wigner and Huntington predicted in 1935 \cite{WignerHuntington} that solid \hm could be metallized at a pressure $\sim$ 25 GPa. Ashcroft predicted in 1968 \cite{Ashcroft1968} that such a metallic hydrogen will be a room temperature BCS superconductor, in view of a large Debye temperature. Unfortunately both predictions have not been confirmed experimentally. Solid \hm refuses to metallize even at a few hundred GPa. Instead, it undergoes a series of structural changes, where covalent bonds survive and get reorganized; it remains insulating. In view of this, interest in the community shifted to hydrogen rich solids. Silane, SiH$_4$, has yielded under pressure and becomes superconducting with a modest Tc $\sim$ 20 K, as found by Eremets et al \cite{EremetsSilane}.  Interestingly, in a recent paper Eremets et al., also report superconductivity \cite{EremetsH2S} in another hydrogen rich solid, \hsm, and a much higher Tc $\sim$ 203 K at a pressure of 200 GPa .

These exciting results needs to be reconfirmed by measurement of Meissner effect, Josephson tunnelling etc. Theory papers have appeared, based on ab-initio calculation, before and after the experiment. They predict a variety of structures and high Tc superconductivity, based on electron-phonon interaction mechanism. Unfortunately, the structure of \hsm is not known experimentally. In an earlier (2004)  experiment X Ray scattering revealed a structure at high pressures. There is molecular dissociation and formation of sulphur-sulphur bond and sulphur helices, known in allotropes of S, Se and Te. Position of hydrogen atom could not be determined, in view of a low electron number associated with H atom. In the absence of phase separation, H atoms or H molecules are likely to occupy small interstitial positions in the densely packed sulphur helix lattice. Sulphur atom has a large atomic radius compared to H atom.

In our theory \cite{GBH2S} we suggest an organization principle, where covalent bonds continue to survive, but may change their spatial pattern. This arises from the kinetic energy gain in every covalent bond by the electron pair being in an orbitally symmetric state. That is, two electrons behave like bosons occupying the same quantum state (albeit with some correlation hole).  Antisymmetry is taken care of by the singlet spin state. From phenomenology and from study of available structures in theory and experiments in \hm and \hsm, I find a strong tendency for total covalent bond number conservation.  There is a great resistance, just from energetics, to form simple unfilled bands and fermi sea, which is a state with liberated single electrons.

 I view these paired valence electrons that form covalent bonds in molecules as stable but confined Cooper pairs. Thus in \hsm the two saturated covalent bonds correspond to two \textit{confined Cooper pairs}. Molecular solid \hm and \hsm are in this sense \textit{Cooper pair insulators}. They are so deeply bound that this is irrelevant in normal situations. However, high pressure, a peculiar liberator,  could deconfine some of these confined Cooper pairs and create a superconducting state. That is pressure, under some conditions and for a range or pressures, prefer to liberate paired electron rather than single electrons.
 
Specifically I suggest the following structural and valence bond reorganization for the superconducting \hsm at high pressures. It leads to strong sulphur-sulphur bond and helical chains or a structure with a dominant saturated S-S bonds. Some dissociated H atoms can remain neutral H-atoms and some as \hm molecules.  At high pressure strongly bonded S atoms form a dense packing. Because of its large atomic radius S atoms leave only small interstitial space for H atom or \hm molecule. The interstitial crystalline network which accommodates H atoms, can be quasi-1, quasi-2 dimensional or 3 dimensional. The H-H distance and H sublattice structure is now dictated by the sulphur sublattice.  In general the structures do not allow for saturated H-H bonds. For example a dimerized H-atom chain allows for saturated H-H bonds. Whereas, uniform chain or a honeycomb lattice type structure leads to valence bond resonance. I find that in some pressure ranges, the H-H transfer matrix elements (direct and through sulphur atoms) is small compared to ionization energy of H-atom, leading to possibility of a Mott insulating sublattice of neutral H atoms. 
 
 In my picture, the Mott insulating subsystem of neutral H-atoms is in general unstable to internal doping. Because, in general, a differing electro negativity of the sulphur and H sublattices will result in a charge transfer between the two subsystems. This charge transfer dopes the H atom Mott insulator and opens a door for superconductivity in a doped Mott insulator. My estimates of the doped Mott insulator parameters, for a few structures for \hsm available in the literature from LDA calculation gives possibility of superconductivity reaching the scale of 200 K, as seen in the experiment. 

Pressure induced dissociation in \hsm has been suggested to create H$_3$S, following a phase separation. Our picture goes through for such hydrogen rich solids, including a recent pressure induced superconductivity in PH$_3$ \cite{EremetsPH3} with Tc exceeding 100 K.

\vskip 0.75 cm
\noindent {\bf{\large 11. Acknowledgement}}
\vskip 0.5 cm

Interaction and collaboration with Anderson, since 1983 has been inspiring and wonderful. It is indeed amazing that one man could do so much to science. It is equally amazing that one man could influence so much, those who cross his path.  It gives me great pleasure to congratulate Anderson on this happy occasion and wish him, Joyce and Susan all well. I thank DAE (India) for the Raja Ramanna Fellowship and the Science and Engineering Research Board (SERB, India) for the SERB Distinguished Fellowship. Additional support was provided by the Perimeter Institute for Theoretical Physics. Research at Perimeter
Institute is supported by the Government of Canada through the Department of Innovation, Science and Economic Development Canada and by the Province of Ontario through the Ministry of Research, Innovation and Science.


\begin{references}
\bibitem{AndersonGardenPicture} When I was looking at the sketch at his office wall years ago, Anderson told me that the sketch was by Robert Dynes, then a colleague at Bell Labs. When I contacted Bob Dynes recently, he regrets not having such a good artistic ability ! Piers Coleman has sent some queries to find the name.
\end{references}
\end{document}